\documentclass[11pt]{article}

\usepackage{booktabs}
\usepackage{dcolumn} 
\usepackage{epstopdf}
\usepackage{fourier}
\usepackage{fullpage}
\usepackage{graphicx}
\usepackage{hyperref}
\usepackage{longtable} 
\usepackage{natbib}
\usepackage{rotating}
\usepackage{tabularx}
\usepackage{amsmath}
\usepackage{algorithmic} 
\usepackage{algorithm2e}

\hypersetup{
  colorlinks = TRUE,
  citecolor=blue,
  linkcolor=red,
  urlcolor=black
}

\begin{document} 

\title{The Tragedy of Your Upstairs Neighbors: \\ Is the Airbnb Negative Externality Internalized?}

\date{\today}

\author{John J. Horton \\ NYU Stern\footnote{Author contact information, datasets and code are currently or will be available at \href{http://www.john-joseph-horton.com/}{http://www.john-joseph-horton.com/}. Thanks to Andrey Fradkin for helpful comments.} }
\maketitle

\begin{abstract}
\noindent A commonly expressed concern about the rise of the peer-to-peer rental market Airbnb is that hosts---those renting out their properties---impose costs on their unwitting neighbors.   
I consider the question of whether apartment building owners will, in a competitive rental market, set a building-specific Airbnb hosting policy that is socially efficient. 
I find that if tenants can sort across apartments based on the owners policy then the equilibrium fraction of buildings allowing Airbnb listing would be socially efficient. 
\end{abstract} 

\section{Introduction}
The peer-to-peer accommodation rental platform Airbnb has grown rapidly: 
since its founding in 2008, over 25 million guests have used the service and there are over 1 million properties listed on the marketplace.\footnote{Airbnb.com/about/, accessed online February 1st, 2015.}
Airbnb, along with other so-called computer-mediated platform companies have bedeviled policy-makers trying to adapt existing regulations to a phenomena that was not foreseen when these regulations were developed.    

The social benefits of Airbnb are clear enough---underutilized resources are put to use.
However, there are several policy-relevant critiques.
Perhaps the most substantive critique is that Airbnb allows ``hosts'' (those renting out properties) to impose a cost on their neighbors, particularly in apartment buildings.  
If Airbnb hosts bring in loud or disreputable guests but, critically, still collect payment, then it would seem to create a classic case of un-internalized externalities that existing illegal hotel laws are intended to prevent: the host gets the money and her neighbors get the noise. 
This regulatory arbitrage is a critique of both Airbnb and so-called ``sharing economy'' companies generally \citep{malhotra}.

For private homes, this issue is more or less non-existent, but in apartment buildings where individuals live in close proximity and there are often heavily used common spaces, the issue is real. 
There is at least some reason to believe that Airbnb guests with no stake in maintaining good relations with neighbors will be more troublesome than the hosts usually living in a location. 
Dean Baker in a recent Guardian column argued: 
\begin{quote}
``[The neighbors of Airbnb hosts] in
condo, co-ops or apartment buildings may think they have the right not
to be living next door to a hotel (which is one reason that cities
have zoning restrictions).''\footnote{The Guardian, ``Don't buy the `sharing economy' hype: Airbnb and Uber are facilitating rip-offs'', May 27th, 2014. Accessed online at \href{http://www.theguardian.com/commentisfree/2014/may/27/airbnb-uber-taxes-regulation}{http://www.theguardian.com/commentisfree/2014/may/27/airbnb-uber-taxes-regulation}} 
\end{quote}
Unlike concerns about tax collection for Airbnb---which public policy can remedy fairly easily---a business built on beggar-thy-neighbor is very unattractive.

Scenarios with private benefits and social costs commonly arise when setting public policy. 
The vivid metaphor used to describe these scenarios is the ``tragedy of commons, '' in which individuals making individually rational decisions about how much of a resource to consume lower overall social welfare.  
Similarly, in the Airbnb context, because the individual host does not compensate or consider the interests of their neighbor (I assume that individual tenants are not able to compensate each other directly through Coasian bargaining \citep{coase1960problem}), there is potentially ``too much'' hosting. 
Of course, the ``too much'' depends on the size of the private benefit from hosting and the size of the social costs. 

The tragedy of the commons analogy to Airbnb is not quite right, as an apartment building is not in a state of nature or held in commons: the owners of the building can decide whether to allow their tenants to list on Airbnb. 
Given that owners can decide a policy for their building, what will happen? 
Ideally, researchers could be empirical about this question, but data would difficult to come by, especially since Airbnb is growing rapidly and the long-term equilibrium is unknown.\footnote{There is some work now trying to quantify the effects of Airbnb on the hotel industry, which is a question more amenable to empirical analysis than the question considered here \citep{zervas2014rise}.} 
To help analyze the issue, I consider a simple model in which there are private benefits and un-internalized social costs from Airbnb listing. 

Apartment building owners decide whether to allow their tenants to list on Airbnb. 
Tenants sort, with those that want to list on Airbnb at the market price moving to those apartment buildings that allow it and those that do not moving to apartment buildings were listing is forbidden.  
The model is similar in motivation to \cite{tiebout1956pure}, where the apartment owners decision is analogous to the choices local governments make about what public goods and taxes to offer. 
In the Tiebout model, preferences are exogenous, with different people have difference preferences over public goods. 
However, in the model I consider, the private benefit of being an Airbnb host depends on the market price of being a host, which in turn depends upon the number of apartment owners that allow listing on Airbnb. 

Because I assume there are no added costs to apartment owners from their choice of Airbnb policy, in a competitive market rents are equalized across apartment buildings. 
When this is the case, then no tenant wants to switch apartment buildings, meaning that everyone who wants to list on Airbnb can and those that do not are in a building where it is not allowed. 
For an Airbnb-listing host in a building that allows listing, to be just in different between building types, the benefits from listing must be exactly equal to the costs from everyone else in the building listing on Airbnb. 
If the benefits were higher---meaning that the host/tenant was getting a surplus---then the building owner could charge a higher rent, capturing some of their tenants surplus. 
This would in turn make being an Airbnb-friendly apartment owner more attractive than the alternative and would bring more owners in the Airbnb-friendly camp.

This process would stop once tenants were no longer getting any surplus, which occurs precisely when the benefit from being an Airbnb host equals the cost of everyone else in the building being an Airbnb host. 
This characterization---benefits equal to costs---is precisely the same condition that would make the presence of Airbnb efficient: trades would occur up until the marginal benefit equals the marginal cost.  
  
\section{A model of owner-decision making and the market for Airbnb rentals}
There are a total of $A$ apartment buildings in some market, each with $n$ tenants.  
The apartment gives each tenant a net utility of $u_0$. 
The fraction of people that would list their apartments on Airbnb is $f(p)$, where $p$ is the market price for an Airbnb unit, with $f'(p) > 0$. 
For now, we will assume that the $f(p)$ who would host and the $1-f(p)$ who would not are mixed across apartment buildings. 

The private benefit to a host from listing is $p$. 
We will assume that hosts are fully capturing the social benefit of the listing (i.e., we assume that the Airbnb users they are renting to are just indifferent). 
Let us assume that it generates a cost to other tenants of $c$. 
We will assume that all tenants (including the host) bear the cost and so total costs are $cn$.
It would be socially optimal for a host to list on Airbnb if 
\begin{equation} \label{eq:optimal}
p = cn,
\end{equation} 
or the private benefit equaled the social cost. 
For an individual tenant, the total costs of being in an apartment where Airbnb rentals are allowed is in expectation $cnf(p)$.
Those tenants that do not list obviously get no benefit.  

Let $D(p)$ be the demand for Airbnb rentals. 
First let us assume that building owners exert no control and individuals decide whether to become hosts. 
In equilibrium, for supply to meet demand,  
\begin{equation} 
D(p) = A n f(p).
\end{equation} 
In this equilibrium, $p$ is set strictly by demand for Airbnb and $f(\cdot)$.  
The social cost of renting is not internalized and if $cn > p$, then the social costs of Airbnb exceed the benefits. 
Because every person who wants to be a host can, it is likely that $p$ is low, as there is lots of supply. 
When $p$ is low, it is less likely that the social welfare condition is met. 

Now let us assume that building owners can set a rule for their apartment building. 
They will set a blanket policy and cannot take or make side-payments to/from tenants. 
At first, when all tenant types are ``mixed'' across apartments, any policy imposed by a landlord will leave some tenants happy and others unhappy. 
Over time, they will move into apartment buildings of the right ``type.''\footnote{Rather than full mixing, it seems likely that types are ``correlated'' within buildings, so the sorting would be less than if tenants were full mixed.} 

\begin{figure} 
\caption{Comparing the utility of being an Airbnb host to not being an Airbnb host \label{fig:diagram}}
\centering
\begin{minipage}{0.65 \textwidth}
\includegraphics[width = \linewidth]{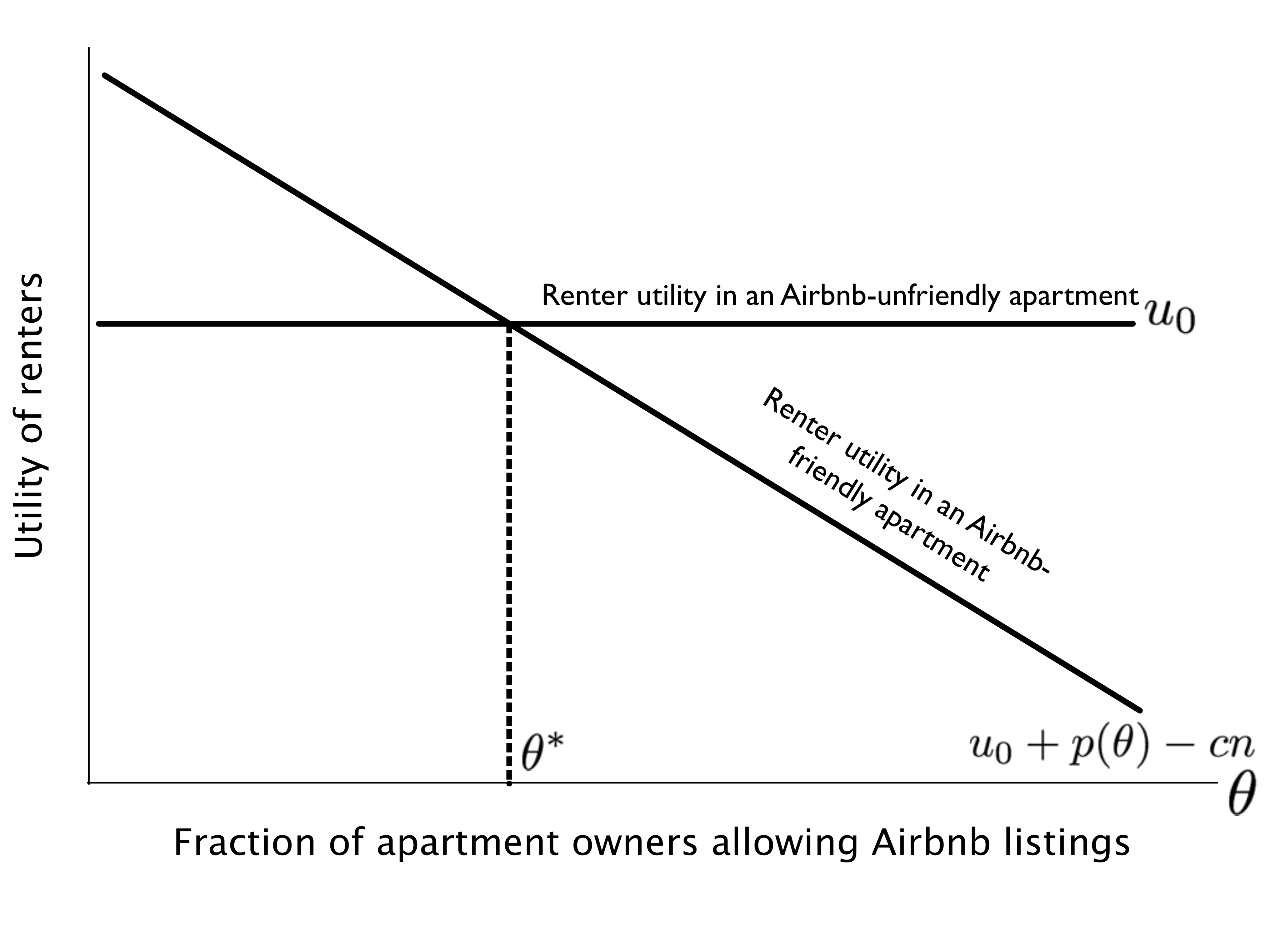}
\end{minipage}
\end{figure}

Figure~\ref{fig:diagram} shows the utility of renters in buildings that allow Airbnb listing (downward sloping curve) and those that do not. 
The x-axis is the fraction of building owners, $\theta$, that allow Airbnb listing. 
For individuals in buildings that do not allow Airbnb listing, their utility is a constant, $u_0$. 
For those that do allow listing, utility is $u_0 + p(\theta) - cn$ (since everyone in the Airbnb-friendly apartment lists), 
where $p(\theta)$ is the inverse demand curve for Airbnb rentals. 

In a competitive equilibrium, a building owner would be indifferent between allowing Airbnb hosting and not, and so rents would be equal in either kind of building.
For rents to be equal, tenants in an apartment that allows hosting has to be indifferent between the two apartment types, meaning that 
\begin{equation}
u_0 + p(\theta^*) - cn = u_0 
\end{equation} 
which implies that $p(\theta^*) = cn$. 
We know this fraction $\theta^*$ exists because $p'(\theta) < 0$, since demand curves slope down. 
This is the social optimal condition from Equation~\ref{eq:optimal}.  

\section{Discussion} 
If $f(p)$ is fairly flat, i.e., supply is fairly inelastic, then changes in price are not likely to require much sorting to ``fix.''
The costs of sorting are not captured and they could be substantial. 
Given that most people are loss averse---valuing a gain less than an equivalently sized lost---it seems likely that there would inefficiently low number of building owners allowing Airbnb.  
For landlords renting below the market price---such as those with rent stabilized apartments---or those in charge of public housing policy---they would presumably be influenced by whatever direct costs hosting generates, if any.
Given that the main direct cost would probably be complaints, they too would be likely to not allow Airbnb listing.  

\bibliographystyle{aer}
\bibliography{airbnb.bib}

\end{document}